# No Women No Innovation? The Effect of Women on Boards on Hard and Soft Innovation in SMEs

*Francesca Pascale*
Scuola Superiore Sant'Anna
Institute of Management
Piazza Martiri della Libertà, 33,
56127 - Pisa, Italy
E-mail: francesca.pascale@santannapisa.it
ORCID: 0009-0007-2998-5834

*Saverio Barabuffi**
Scuola Superiore Sant'Anna
Institute of Management and L'EMbeDS Department
Piazza Martiri della Libertà, 33,
56127 - Pisa, Italy
E-mail: saverio.barabuffi@santannapisa.it
ORCID: 0009-0005-4619-9152
*corresponding author

*Giulio Ferrigno*
Scuola Superiore Sant'Anna
Institute of Management and L'EMbeDS Department
Piazza Martiri della Libertà, 33,
56127 - Pisa, Italy
E-mail: giulio.ferrigno@santannapisa.it
ORCID: 0000-0003-4128-2190

# No Women No Innovation? The Effect of Women on Boards on Hard and Soft Innovation in SMEs

**Abstract**

The role of women on corporate boards and its impact on innovation remains heavily debated. While existing literature offers conflicting perspectives on whether female board representation spurs or hinders innovation, most research has focused exclusively on large firms. This study addresses this gap by investigating the causal impact of women on boards on both "hard" (technological) and "soft" (non-technological) innovation within small and medium-sized enterprises (SMEs).

Using a longitudinal dataset of 2,762 Italian innovative SMEs from 2015 to 2024 and a 2SLS identification strategy, we find highly nuanced patterns. For hard innovation, the instrumented model reveals a robust positive causal impact, reversing a negative baseline coefficient and exposing a severe negative selection bias. For soft innovation, a stable positive impact is confirmed. Our findings provide crucial causal evidence for SME governance, while offering tailored insights for policymakers and practitioners to incentivize women' management roles in SMEs.

**Keywords**: Women on Board; Gender; SMEs; Patents; Trademarks

**JEL codes:** J16; L25; O31

**Disclosure statement**

No potential conflict of interest was reported by the author(s).

## 1. Introduction

In recent decades, academics and practitioners have strongly debated the role of women on boards, attracting increasing attention of policy makers (Smulowitz et al., 2025). Several countries (i.e., Spain, Italy, Norway, California and many others) have consequently introduced laws to guarantee a more balanced and fair composition of corporate boards (Greene et al., 2020). Besides social equality consequences, these political interventions also positively influence the value maximization of the firm. Value maximization literature states that more women on boards positively affect firm value through a better decision-making process (Griffin et al., 2021), influencing monitoring and resource provision (Hillman and Dalziel, 2003). Nevertheless, frequently women are placed into board positions due to tokenism (King et al., 2010), without actually influencing firm's management, or they are welcomed on board during crisis periods and high-risk situations, commonly called "glass cliff" situation (Ryan and Haslam, 2005, 2007; Glass and Cook, 2020). Additionally, female board members often perceive a lower status, and they are less valued as a group member (Duguid et al., 2012; Vial et al., 2016).

Existing literature presents conflicting perspectives regarding the influence of women board membership on innovation outputs. Some studies state that women are more risk-averse negatively influencing the level of innovation in the firm (e.g., Nelson, 2015; Faccio et al., 2016). Empirical research has shown that on average women choose options with fewer financial risks than male counterparts (Sapienza et al., 2009). Additionally, women often see their position as more precarious than the one of their male counterparts, amplifying their risk-averse approach (Smith et al., 2019; Glass and Cook, 2020)

Other works provide opposite results mainly focusing on the improvement of decision-making process given by the presence of women on the board that results in more innovation (Adams and Ferreira, 2009; Cumming et al., 2015; Griffin et al., 2021; Hemmert et al., 2022; Smulowitz and Smulowitz, 2024). Heterogeneity in boards spurs broader perspectives which allow the management to be involved in in-depth conversations (Watson et al., 1998; Rao and Tilt, 2016) and recombine different ideas that lead to new products, services, or marketing methods (Dai et al., 2019; Grekou et al., 2023).

However, past research focused on the role of women on corporate boards in large firms in relation to firm innovation (Smulowitz et al., 2025), there remains a gap in understanding how these dynamics works for SMEs. This study aims to extend empirical investigation by exploring whether similar effects hold for SMEs, contributing to the broader understanding of the effect of gender diversity in corporate governance on innovation within smaller enterprises. We analyse innovation in terms of hard and soft innovation. Specifically, "hard" innovation refers to all the technological innovations which are tangible and science-based, often resulting from significant R&D expenditures (Block et al., 2025), while "soft" innovation indicates all the non-technological innovations (e.g., in product, service, marketing, or business models) (Flikkema et al., 2019). Hence, studying both hard and soft innovation can be a fundamental source of information of SMEs innovation, especially in those presenting gender diversity.

As such, we aim to investigate the effect of women on board on hard and soft innovation in SMEs, contributing to filling the gap in the literature and answering the following research question: *what is the effect of women on boards on hard and soft innovation in SMEs?*

To address this gap, we focus on the Italian context as Italy is characterized by a widespread SMEs ecosystem. Specifically, about 80% of the Italian workforce is employed in SMEs (ISTAT, 2023), qualifying Italy as a strong empirical setting to validate our hypotheses and provide significant contributions to the literature (Cucino et al., 2024). To investigate the temporal dynamics of the role of women on board on the innovative performance of SMEs, this study utilizes a comprehensive longitudinal dataset of 2762 SMEs from 2015 to 2024 which are registered as "innovative SMEs" in a specific section of the Italian Chamber of Commerce's register, demonstrating their status through objective indicators. We implement a Two-Stage Least Squares (2SLS) identification strategy, leveraging exogenous variations in local female labour supply driven by sub-national childcare infrastructure density. Our empirical findings reveal highly nuanced, heterogeneous patterns across different dimensions of SMEs innovative activity, offering deep insights into the underlying nature of firm-level endogeneity. For hard innovation, we observe a sign reversal, since the baseline estimation yields a negative coefficient, whereas the instrumented 2SLS model uncovers a robust, statistically significant positive causal impact. This sign reversal is symptomatic of severe negative selection bias, where underperforming firms systematically appoint female leaders, masking the true asset value of gender diversity. In contrast, when examining soft innovation, the OLS and 2SLS coefficients retain the same positive sign, with the instrumented effect exhibiting increased statistical magnitude. This stability indicates that for this specific innovative dimension, the baseline model is not confounded by negative sorting, but rather by standard attenuation bias or unobserved positive firm characteristics.

Our work contributes to the previous studies in several ways. First, it contributes by filling the gap in the literature, investigating the relationship between women board representation and innovation within SMEs (Arcuri et al., 2023). Second, we examine the effect of women board representation on both hard and soft innovation (De Vries et al., 2017; Flikkema et al., 2025). Finally, this paper provides causal evidence regarding how the inclusion of women in SME management directly drives the extensive margin of firm innovation.

The paper also offers managerial contributions, suggesting managers and founders be more flexible and open-minded, and policy contributions, highlighting the need for targeted policy interventions moving beyond general innovation subsidies. Policymakers should introduce tailored incentives to support women-led firms and intervene to increase patenting activity among women-led firms.

The rest of the paper is organized as follows. In Section 2 we review literature on gender diversity among SMEs, and we develop our hypotheses focusing on the influence of women on board on hard and soft innovation. In Sections 3 we describe the data and methodology adopted and in Section 4 we discuss our results. In Section 5, we highlight implications for theory, policy, and practice and we present possible limitations of the study suggesting some avenues for future research.

## 2. Theoretical Background and Research Hypotheses

### 2.1. Gender Diversity on Boards

The relationship between female representation in management roles and firm outcomes has been extensively examined through multiple theoretical lenses, particularly within the value maximization and corporate governance literature (Smulowitz et al., 2025). In this context, the presence of Women on Boards, and, consequently of gender diversity is often associated with improved governance mechanism, as it enhances both monitoring and the provision of critical resources (Hillman and Dalziel, 2003). Board composition also shapes firm's strategic cognition and the interpretation of complex opportunities, which is an issue that becomes significantly salient for innovation-related choices (Bauweraerts et al., 2022). Building on this perspective, the inclusion of women contributes to higher-quality strategic decision by influencing the effectiveness of decision-making processes, supporting firm value (Griffin et al., 2021), and firm performance (Smulowitz et al., 2025).

Despite these potential benefits, the actual influence of women directors is not always fully realized mainly due to the presence of tokenism (King et al., 2010; Ain et al., 2022; Childress et al., 2024), whereby women are appointed to boards primarily to signal diversity rather than to actively participate in decision-making. This issue is further amplified by the tendency to appoint women in precarious contexts, such as during crisis periods and high-risk situations, commonly described as "glass cliff" phenomenon (Ryan and Haslam, 2005, 2007; Kulich and Ryan, 2017).

While for large firms there are several studies trying to understand whether female leadership has a positive or negative effect on innovation performance (Smulowitz et al, 2025), very few studies focus on this relationship for SMEs (Arcuri et al., 2023; Mari et al., 2024). There may be significant differences in the SMEs context arising from the consideration that organizational cultures may play a stronger role, spurring a more open and flexible context, which may spur creativity, or a more rigid or hierarchical environment, which may hinder innovation (Tian et al., 2018; Ruiz et al., 2023). Regarding SMEs, some studies focus on the effect of gender diversity on innovation performance in general (Ruiz-Jiménez et al., 2016; Bauweraerts et al., 2022; Veltri et al., 2025; Kiessling et al., 2026), but an important gap remains in analysing the potential different effects of female managers on SMEs' hard and soft innovation performance.

### 2.2. Women on Board and Hard Innovation

Most studies focus on the investigation of hard innovation, usually in terms of patent analysis (e.g., Liu and Zhu, 2024; Smulowitz et al., 2025). Patents capture firms' ability to develop innovations and their total technology portfolio (Ahuja and Lampert, 2001; Berrone et al., 2013), and they are frequently used as a measure of firm innovation performance (Katila, 2000; Kotha et al., 2011). More specifically, there is a tendency to rely on hard and technological innovation when analysing specific sectors (e.g., Block et al., 2025).

For SMEs hard innovation can be especially important as it can strengthen productivity and quality upgrading, enhance appropriability through formal intellectual property protection and signal technology capability to external investors or partners (Block et al., 2025). Since SMEs typically face resource constraints and the so called "liability of smallness" (e.g., Stinchcombe, 1965; Gimenez-Fernandez et al., 2020), understanding the governance conditions that foster innovation performance is therefore particularly important.

In terms of hard innovation, which often requires strict and rigorous processes, debate and creativity may be seen as negative for the efficiency of the process. Even if diversity provides various cognitive resources (Fiedler, 1989; Liu, 2021), these are highly context dependent (Sundriyal et al., 2024). Women on boards may influence hard innovation by broadening the board's cognitive and experiential resources, bringing deeper debate for decision-making. However, since hard innovation requires disciplined routines, stringent evaluation, and coordination across specialized functions, gender diversity may increase task-related disagreement and risk aversion (Sapienza et al., 2009), which may lead to more cautious strategic decisions and, consequently, can slow resource allocation to R&D and reduce perceived process efficiency (e.g., Nelson, 2015; Faccio et al., 2016). SMEs seem to be more rigid and context dependent than large firms, potentially negatively affecting diversity perception in the specific field of hard innovation (Sundriyal et al., 2024) and increase gendered "inequality approach" that can restrict women's access to innovation-relevant resources and legitimacy (Owalla et al., 2021; Chowdhury et al., 2023; Kiessling et al., 2026).

Consequently, gender diverse SMEs face potentially stronger constraints on hard innovations (Ruiz-Palomo et al., 2024) due to women perception of occupying more fragile positions within their organizations (Vial et al., 2016; Smith et al., 2019; Glass and Cook, 2020). In close and traditionalist contexts, like the one of hard innovation, firms may not be able to fully exploit the potential of diverse cognitive resources (Cetrulo et al., 2020). Consequently, diversity provokes stronger debate consuming more time and resources to take decisions, potentially hindering the efficiency of the process (Van Knippenberg et al., 2004; Rao and Tilt, 2016).

As such, we propose the following hypothesis:

H1: The *presence of Women on Board negatively influences hard innovation in SMEs*

### 2.3. Women on Board and Soft Innovation

Within SMEs, innovation is typically understood not only in terms of patents or R&D activities, but also as the ability to adapt products, and services (Kiessling et al., 2026). Unlike "hard" innovation, soft innovation primarily reflects intangible and market-facing forms of novelty that are often not protected through patents. Incorporating soft innovation into the measurement of innovation can substantially improve the identification of innovation activity among SMEs, for which innovation frequently takes incremental, customer-driven and commercialization-oriented forms (Morales et al., 2024; Block et al., 2025). This is particularly important because SMEs often compete through differentiation, branding, niche positioning, and rapid adaptation rather than through large-scale R&D investments (Gimenez-Fernandez et al., 2020).

Usually, soft innovation is measured by trademarks in empirical studies (Gotsch and Hipp, 2012; Flikkema et al., 2014, 2019; Block et al., 2025). Trademarks capture firms' investments in branding, product/service differentiation, and market introduction of new offerings, thereby providing a useful proxy for commercialization-oriented innovation outputs. Past studies suggest that trademarks are an indicator of innovation in the service sector (Mendonça et al., 2004; Gotsch and Hipp, 2012), as well as in the creative and cultural industries (Castaldi, 2018). Accordingly, trademarks help operationalize soft innovation in sectors where novelty is expressed through positioning, design, and customer value propositions rather than patentable technologies.

Nevertheless, the emerging insight is that women presence in management positions can strengthen soft innovation by improving strategic decision-making and broadening the firm's interpretation of markets and stakeholders needs (Griffin et al., 2021; Hemmert et al., 2022). Bednar et al. (2021) demonstrate that female-owned firms' projects have greater probability of being commercialized than in male-owned firms. Hence, gender-diverse boards tend to enhance information processing and increase the consideration of alternative strategic options, which can translate into greater experimentation of soft innovation (Smulowitz and Smulowitz, 2024). In this context, diversity can be understood not merely as a workforce feature, but as a tool shaping innovation potential by affecting decision-making processes and signalling openness to heterogeneity (Veltri et al., 2025). Indeed, soft innovation is less dependent on strict technical routines and more reliant on creativity and knowledge recombination (Rao and Tilt, 2016). Consistent with this reasoning, female SMEs entrepreneurs are often more effective in leverage innovation compared to their male counterparts, emphasizing adaptability and flexibility, which are closely aligned with the nature of soft innovation (Arcuri et al., 2023). Also, some studies found that board gender diversity of innovative SMEs positively influences performance outcomes, highlighting the importance of inclusiveness in fostering innovation (Schifilliti and La Rocca, 2024). Furthermore, Mari et al. (2024) suggest that women entrepreneurs often engage in innovation through open innovation, which facilitates knowledge inflows and collaboration and may therefore increase soft innovation outputs such as trademarks and new market offerings.

As such, we develop the following hypothesis:

H2: The *presence of Women on Board positively influences soft innovation in SMEs*

## 3. Data and Methodology

### 3.1 Data and Variables

We build a unique dataset collecting information from the Italian Chamber of Commerce database, BvD Aida database, BvD Orbis database, BvD Orbis Intellectual Property database, and EUIPO TMView database. The Italian Chamber of Commerce database is a national institution responsible for validating all companies' administrative and legal requirements. The law requires that all innovative SMEs must register in a specific section of the Italian Chamber of Commerce's register and provide information regarding board composition. Italian policy makers legally define innovative SMEs by DL 3/2015 and require them to register in a specific section of the Italian Chamber of Commerce. Specifically, Innovative SMEs in Italy must employ fewer than 250 individuals and have either an annual turnover not exceeding €50 million or a total annual balance sheet below €43 million. Within this framework, the regulatory criteria allow for a clear and effective identification of innovative SMEs.

The innovative nature of these firms is not presumed but must be demonstrated through objective indicators. In particular, an SMEs is defined as "innovative" if it satisfies at least two out of three conditions. First, it must show a significant commitment to innovation, with R&D expenditures amounting to no less than 3% of the higher value between total production costs and total production value of the latest approved financial statements. Second, the firm must rely on highly qualified human capital, either by employing a substantial proportion of PhD holders, doctoral candidates, or alternatively by maintaining a workforce in which a significant share holds advanced academic degrees. Third, the company may demonstrate its innovative capacity through the ownership or licensed use of intellectual property rights.

Additionally, innovative SMEs are required to confirm annually that they continue to meet at least two of the prescribed criteria.

From this dataset, we obtain 2948 innovative SMEs in Italy registered between 2015 and 2025. Moreover, we employ the BvD Aida database to collect financial information. Then, we collect from BvD Orbis the information on firms' management, such as the role, date of appointment and resignation, and gender. Finally, we collect patent data employing Orbis IP, and we manually search trademark data employing EUIPO TMView. By merging data from these several sources, we are able to conduct our empirical analysis on an unbalanced panel of 2762 SMEs for the period 2015-2024[1].

[INSERT FIGURE 1 HERE]

*Dependent variables*

We proxied the dependent variable, *hard innovation,* by relying on patent data (Liu and Zhu, 2024; Smulowitz *et al.,* 2025). Specifically, we employed a binary variable indicating whether a firm filed a patent at time *t* given that SMEs typically patent fewer innovations than larger firms due to scale-related and cost barriers (Athreye *et al.*, 2021). Furthermore, our analysis distinguishes between patent applications and granted patents to provide a more granular view of the role of women in management on firm-level patenting performance.

Then, following Block et al. (2025), we used trademarks as a proxy for *soft innovation*, and we operationalized it as a binary variable which indicate whether a firm filed a trademark at time *t*. To offer a more nuanced perspective, we further disaggregated this measure by distinguishing between trademark applications and registered trademarks.

*Independent variable*

To capture the presence of female leadership within the firm, we construct a time-varying binary indicator, $WoB_{i,t}$, which takes the value 1 if firm *i* employs at least one woman in a top or middle management position[2] in year *t*. We intentionally utilize a dummy specification rather than a continuous share to explicitly evaluate the extensive margin of female executive representation. This choice is strongly motivated by the institutional characteristics of the Italian corporate landscape, where women have historically faced systemic glass ceilings and significantly low rates of career advancement into senior leadership roles (Rinaldi and Tagliazucchi, 2021; 2025; Gabbuti and Licini, 2025). By defining our explanatory variable around this critical organizational threshold by moving from an entirely male-dominated management team to one with female representation, we isolate the structural transformation of the firm's leadership dynamics.

[1] In the appendix is reported the firms' distribution for each NUTS 3 region.

In order to create this variable, we leveraged BvD Orbis Database information on firms' management. We retrieved the role, date of appointment and resignation, and gender of top and middle management in 2762 SMEs, hence for each firm we are able to report the management composition in terms of gender. While this firm-level metric accurately captures the presence of women in management position, corporate components are rarely randomly assigned. Potential issues of reverse causality and omitted variable bias necessitate an instrumental variable approach, which we outline in Section 3.1

*Control variables*

Then, in the analysis we employ the following firm-level control variables[3]: i) *Age*, defined in terms of number of years after constitution; ii) *Capitalization*, measured as the ratio of equity to total assets; iii) *Labor intensity,* computed as the ratio of wages to total assets; iv) *Marginality*, proxied as the ratio of revenues to production costs; v) *Asset structure*, measured as the ratio of total assets to fixed assets; vi) *Liquidity,* proxied as the ratio of working capital to total assets; and, vii) *Profitability*, measured as the ratio of profit/loss to total assets.

[2] In the Appendix are reported the description of both top and middle management positions.
[3] We scale the economic variable to account for the size of the firm and we transform them by using inverse sine hyperbolic transformation.

### 3.1 Methodology

To test our hypotheses, we estimated the following linear probability model:

$$Y_{i,t} = \beta_1 WoB_{i,t-1} + X'_{i,t-1}\beta_k + \eta_i + \iota_t + \varepsilon_{i,t}$$

where $Y_{i,t}$ is a dummy variable that measures whether a firm *i* has developed a *hard* or *soft innovation* at time *t,* then $WoB_{i,t-1}$ accounts for the presence of women in management roles of firm *i* at time *t-1,* while $X'_{i,t-1}$ represents a vector of lagged firm-level control variables. Lastly, $\eta_i$, $\iota_t$ and $\varepsilon_{i,t}$ account for firm fixed effects, time fixed effects, and the idiosyncratic error, respectively.

However, to estimate the causal impact of *WoB* on SME innovation, we must address the endogeneity concerns related to biased coefficients if highly innovative firms systematically attract more female talent by implementing a 2SLS identification strategy. We isolate exogenous variation in the local supply of female labor by constructing an instrument based on provincial level childcare infrastructure. Therefore, we adopt a Revealed Comparative Advantage (*RCA*) index (Balassa, 1965) to sub-national kindergarten density, which scales the supply of physical school of the volume of children between 3 and 36 months enrolled in childcare schooling within NUTS 3 region *r* against the aggregate national enrollment baseline for that given year. We build the *RCA* as the Childcare Location Quotient (*CLQ*) as follows:

$$CLQ_{r,t} = \frac{Number\ of\ Kindergaten\ _{r,t}/Enrolled\ Children_{r,t}}{Total\ National\ Kindergarten_t/Total\ National\ Enrolled\ Children_t}$$

if $CLQ_{r,t}$value exceeds unity signifies a higher relative density of sub-national institutional capacity and childcare infrastructure deployment relative to its national peers. By utilizing this relative density index, we effectively purge the institutional indicator of confounding country-wide macroeconomic cycles or uniform regional policy mandates

Then, we introduce idiosyncratic, firm-level variation by interacting the time-varying $CLQ_{r,t}$ with a time varying, firm-specific variable calculated as the firm's total employee size

$$IV\ CLQ_{i,t} = CLQ_{r,t} \times Size_{i,t}$$

The relevance condition rests on the microeconomic premise that robust NUTS 3 regional childcare infrastructure systematically reduces the career-interruption penalties and domestic frictions traditionally faced by high-skilled women, thereby expanding the local supply of qualified female management available to the firm (Bauernschuster and Schlotter, 2015; Yamaguchi et al., 2018). Despite sub-national administrative coordination and deployment of physical school directly expands the female executive talent pool, the exclusion restriction is satisfied through the lack of plausible direct influence of provision of schools to early childhood education on an individual SME's patenting or trademarking through any channel other than the presence of women in the firm's leadership.

Table 1 shows the frequency distribution and descriptive statistics, while Table 2 reports the correlation matrix.

*[INSERT TABLE 1 HERE]*

*[INSERT TABLE 2 HERE]*

## 4. Results

Table 3 reports the results of the linear probability model with fixed effects and the 2SLS in which we estimated the effect of WoB on *hard innovation*. In column 1 we find a negative and non-statistically significant effect of $WoB$ on *patent application* ($\beta = -0.0083, p - value > 0.1$), while column 2 shows a negative and statistically significant results of $WoB$ on *granted patents* ($\beta = -0.0157,\ p - value < 0.1$). However, our estimation may be biased due to endogeneity of the independent variable $WoB$. Hence, we apply a 2SLS estimation to correct for endogenous selection bias. Column 3 shows the 2SLS first stage estimation, in which $IV\ CLQ$ shows a positive and statistically significant effect on $WoB$. Columns 4 and 5 report the 2SLS second stage estimation on *patent application* and *granted patents* in which we find a positive and statistically significant result ($\beta = -0.765,\ p - value < 0.05; \beta = 0.368,\ p - value < 0.05$).

While our baseline hypotheses hinted at a negative correlation, the 2SLS second-stage estimates, by correcting for severe endogenous selection, yield a positive and statistically significant coefficients thus rejecting H1. Specifically, our baseline models were confounded by negative selection, such as glass cliff dynamics (Glass and Cook, 2020), where underperforming or high-risk SMEs systematically appoint women in management positions (Ryan and Haslam, 2005, 2007). This divergence is by our positive first-stage relationship, which confirms that a higher sub-national CLQ increases the probability of an SME appointing female managers. This configuration reveals that correlation models suffer from severe negative selection bias, driven by unobserved domestic time-allocation constraints and "double-shift" penalties.

In environments with scarce infrastructure, women are often only able to attain executive roles within underperforming or highly volatile SMEs, or under extreme personal coordination friction that compromises their strategic bandwidth (Kulich and Ryan, 2017), hence biasing the OLS coefficient downward. By utilizing our localized institutional instrument, we effectively isolate female managers who are structurally supported by their local infrastructure. Once freed from these spatial and temporal care bottlenecks, the presence of women in management exerts a robust, positive causal impact on the extensive margin of SME *hard innovation* activity.

[INSERT TABLE 3 AROUND HERE]

Table 4 shows the findings of the linear probability model with fixed effects and the 2SLS estimation of the effect of WoB on *soft innovation.* Column 1 and 2 indicates a positive and statistically significant effect of WoB on *trademark application* and *registered trademarks* $(\beta = 0.0142, p - value < 0.05; \beta = 0.0132, p - value < 0.1)$. Column 3 reports the 2SLS first stage estimation.

Column 4 and 5 report the 2SLS second stage estimation on *trademark application* and *registered patents* in which we find a positive and statistically significant result (), thus confirming H2

Since *soft innovation* relies heavily on market responsiveness, consumer behavioral insights, brand repositioning, and organizational agility (Flikkema et al., 2019), SMEs are already actively positioning women in leadership roles where their cognitive diversity can directly translate into market-facing commercial properties (Tian et al., 2018). The IV strategy further proves that women on board can drive *soft innovation* as an intrinsic corporate asset that manifests even in the presence of baseline operational frictions.

[INSERT TABLE 4 AROUND HERE]

### 4.1 Robustness checks

In Table 5 we perform robustness checks by regressing the binary indicator, $\mathrm{WoB}_{\mathrm{top}}$, which takes the value 1 if firm *i* employs at least one woman in a top management position in year *t*.

Column 1 and 2 qualitatively confirms the results of the correlation effect shown in Table 3, while column 3 and 4 support the OLS results of Table 4.

Column 6 and 7 show the results of the 2SLS second stage, and we find a positive and statistically significant effect of $\mathrm{WoB}_{\mathrm{top}}$ on *patent application* and *granted patents* ($\beta = 0.6793, p - value < 0.05; \beta = 0.3267, p - value < 0.05, respectively$) confirming the results presented in Table 3. In column 8 and 9 we report the results of the 2SLS second stage, and we find a positive and statistically significant effect of $\mathrm{WoB}_{\mathrm{top}}$ on *trademark application* and *registered trademarks* ($\beta = 0.4294, p - value < 0.05; \beta = 0.3763, p - value < 0.05, respectively$), thus confirming the result shown in Table 4.

[INSERT TABLE 5 AROUND HERE]

## 5. Conclusions

Our study provides several contributions to the literature on innovation, strategic leadership, and entrepreneurship. First, this research addresses a significant gap in the field by unpacking the nuanced relationship between women on board representation on innovation within SMEs (Bednar et al., 2021; Arcuri et al., 2023; Chowdhury et al., 2023). Specifically, we move beyond aggregate measures of innovation by distinguishing between “hard” technological innovation and “soft” non-technological innovation (Athreye et al., 2021; Smulowitz et al., 2025; Block et al., 2025). By doing so, we provide a potential answer to conflicting theoretical perspectives on whether female leadership promotes risk-aversion or fosters organizational change (Smulowitz et al., 2025). Indeed, only a few studies focus on soft innovation, which is increasingly recognized as essential for organizational survival in competitive environments (Azzam, 2022; Smulowitz et al., 2025).

Second, while previous studies have extensively addressed this relationship in large corporations, we shift the analytical lens to SMEs (Arcuri et al., 2023; Mari et al., 2024). This focus is critical because SMEs do not merely represent "small large firms", but they face unique “liabilities of smallness”, such as resource constraints and different propensities to seek formal intellectual property protection, that fundamentally alter their innovation trajectories (Gimenez-Fernandez et al., 2020; Block et al., 2025). Moreover, this type of firms is fundamental for most of European economic environments (Cucino et al., 2024) and significant differences compared to large firms may emerge due to their specific features (Arcuri et al., 2023). As such, understanding these specific features allows for more effective policy recommendations that move beyond “one size fits all” approaches (Gimenez-Fernandez et al., 2020).

Third, our research provides a methodological advancement by integrating data from the Italian chamber of commerce, Aida-BvD, and Orbis IP. Indeed, this study constructs a unique longitudinal dataset that combines board compositions, financial information, patent and trademark data overcoming the limitations of cross-sectional and survey-based methods databases prevalent in this research streams (Athreye et al., 2021; Ruiz-Palomo et al., 2022; Weseman et al. 2025; Pascale et al., 2025).

Finally, we provide causal evidence regarding the inclusion of women in SME management directly drives the extensive margin of firm innovation. Our instrumental analysis reveals that findings focusing on correlation suffers from structural sorting related to tokenism (King et al., 2010) and glass cliff situation (Ryan and Haslam, 2007).

Our empirical framework translates the mechanics of 2SLS identification strategy into actionable directives for national and sub-national policymakers. The causal increases observed in hard and soft innovations are explicitly driven by SMEs whose transition toward gender-diverse management was structurally enabled by a high sub-national density of childcare facilities. Hence, the scarcity of female executives within the SME sector becomes a supply-side infrastructure constraint. Consequently, we argue that European regional development frameworks must explicitly reclassify early childhood infrastructure from a passive social welfare expenditure to an active catalyst for industrial innovation. By optimizing the local ratio of schools to enrolled children, regional administrations exogenously expand the high-skilled female talent pool and directly lowers the spatial and temporal friction of care, expanding the effective bandwidth of women already in, or entering, management roles. This localized policy lever lowers the structural barriers that prematurely drain qualified female talent away from corporate leadership pipelines, yielding a direct, measurable return on public capital through the acceleration of private-sector intellectual property registration and long-term commercial competitiveness. Furthermore, policymakers should design targeted intellectual property programs specifically tailored to gender-diverse management teams in SMEs by fast-tracking patent examination processes or offering subsidized trademark filing fees for SMEs that actively dismantle corporate glass cliffs, hence maximizing the allocative efficiency of public innovation subsidies.

This study has also several important implications for managers and entrepreneurs seeking to foster innovation and improve corporate governance in SMEs. The positive first-stage relationship confirms that a supportive local infrastructure directly expands the available pool of female managerial talent. More importantly, the subsequent causal increase in patents and trademarks under 2SLS highlights that the true barrier to corporate innovation was never a lack of female capability, but rather the unobserved tax of domestic time-allocation constraints. Cognitive diversity and problem-solving capabilities of female leaders are maximized when organizational or regional frameworks reduce logistical stress. Forward-thinking SMEs should not treat childcare as an external, state-only issue. To fully internalize the innovation premiums documented in this study, managers should proactively implement flexible spatial-temporal workflows, and localized care partnerships. By actively protecting and freeing the cognitive bandwidth of their executive team, firms can indirectly improve their hard and soft innovation performance.

Despite its strong contributions to the academic field, the present study could present some limitations. The study focuses exclusively on the Italian innovative SMEs, which may limit the applicability of findings to other countries with different regulatory environments, cultural attitudes toward gender diversity, and innovation ecosystems. Future research could explore similar dynamics in other contexts to validate the results or offer cross-country analyses. Finally, the study does not accurately capture the actual influence of female managers hence qualitative studies could complement this research by providing a deeper understanding of their role in SMEs.

## Figures

*Figure 1 - Quartile distribution of firms in Italy at NUTS 3 level*

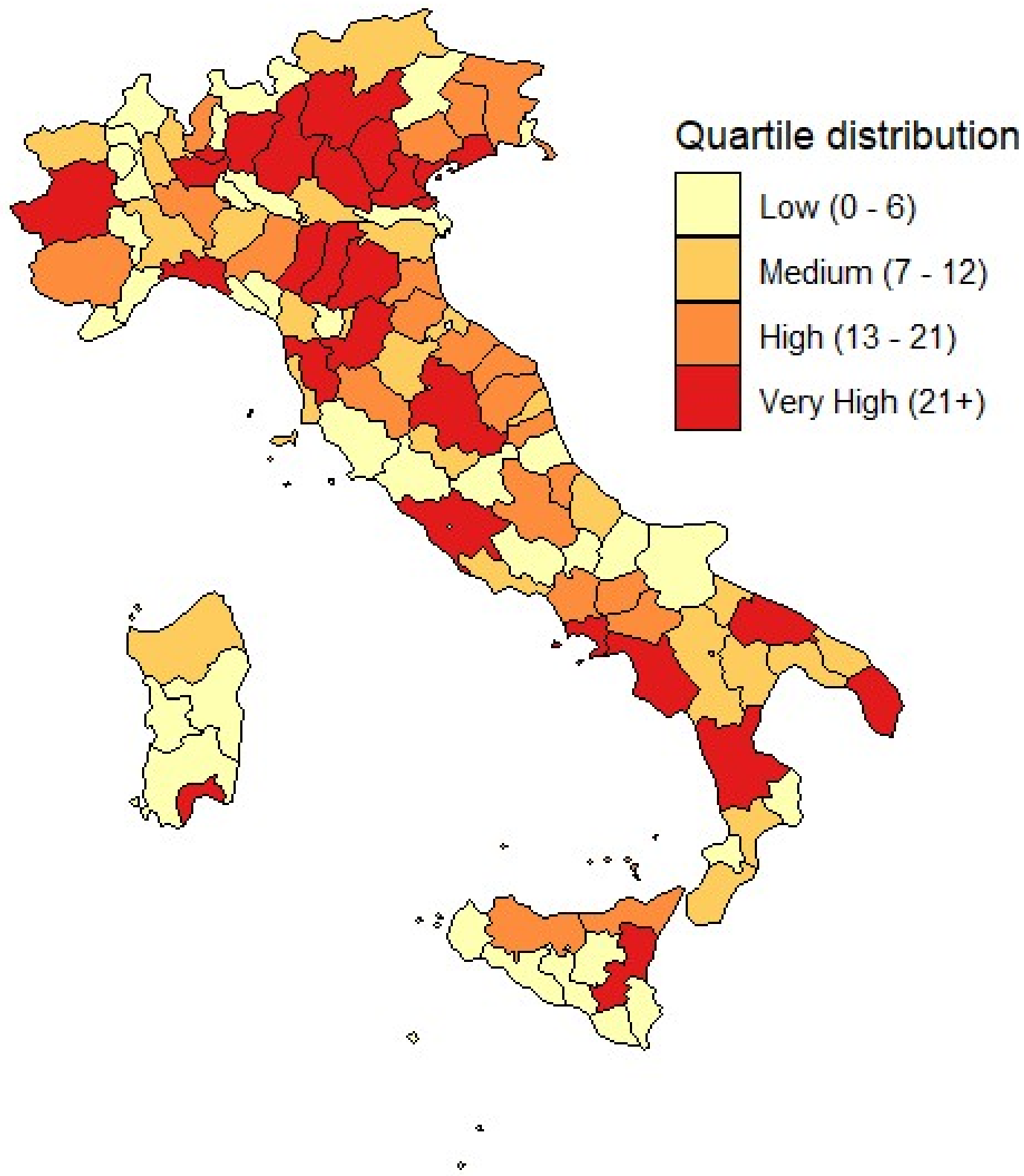

**Tables**

*Table 1 - Frequency distribution and descriptive statistics*

*Frequency distribution*

| | Source | Obs. | 0 | 1 |
|---|---|---|---|---|
| *Patent applications* | BvD Orbis IP database | 24858 | 22058 | 2800 |
| *Granted patents* | BvD Orbis IP database | 24858 | 22911 | 1947 |
| *Trademark applications* | EUIPO TMView | 24858 | 22394 | 2464 |
| *Registerd trademarks* | EUIPO TMView | 24858 | 22449 | 2409 |
| *WoB* | BvD Orbis | 24858 | 20527 | 4331 |

*Descriptive statistics*

| | Source | Obs. | Mean | SD | Min | Max |
|---|---|---|---|---|---|---|
| *IV CLQ* | ISTAT<br>BvD Aida | 20818 | 1.56 | 1.23 | 0.00 | 6.15 |
| *Age* | BvD Aida | 21879 | 2.41 | 1.06 | 0.00 | 5.10 |
| *Capitalization* | BvD Aida | 21286 | 0.36 | 0.28 | -4.74 | 0.88 |
| *Labor intensity* | BvD Aida | 21288 | 0.16 | 0.18 | 0.00 | 3.37 |
| *Marginality* | BvD Aida | 21111 | 0.71 | 0.48 | -0.42 | 12.85 |
| *Asset structure* | BvD Aida | 21073 | 2.29 | 1.16 | 0.88 | 13.25 |
| *Liquidity* | BvD Aida | 21190 | 0.59 | 0.23 | 0.00 | 3.73 |
| *Profitability* | BvD Aida | 21314 | -0.02 | 0.23 | -4.75 | 1.59 |

*Table 2 - Correlation matrix*

| | Patent applications | Granted patents | Trademark applications | Registered trademark | WoB | IV CLQ | Age | Capitalization |
|---|---|---|---|---|---|---|---|---|
| Patent applications | 1.00 | | | | | | | |
| Granted patents | 0.79 | 1.00 | | | | | | |
| Trademark applications | 0.08 | 0.08 | 1.00 | | | | | |
| Registered trademark | 0.08 | 0.08 | 0.98 | 1.00 | | | | |
| WoB | 0.01 | -0.01 | 0.00 | 0.00 | 1.00 | | | |
| IV CLQ | 0.03 | 0.01 | 0.06 | 0.06 | 0.08 | 1.00 | | |
| Age | -0.04 | -0.05 | -0.02 | -0.02 | 0.08 | 0.55 | 1.00 | |
| Capitalization | 0.07 | 0.04 | -0.01 | -0.01 | 0.05 | -0.13 | -0.08 | 1.00 |
| Labor intensity | -0.06 | -0.05 | 0.00 | 0.00 | -0.01 | 0.65 | 0.32 | -0.16 |
| Marginality | -0.09 | -0.06 | 0.02 | 0.02 | -0.03 | 0.23 | 0.23 | -0.02 |
| Asset structure | -0.04 | -0.01 | -0.01 | -0.01 | -0.02 | 0.07 | 0.04 | 0.04 |
| Liquidity | -0.03 | -0.01 | -0.01 | -0.01 | -0.01 | 0.12 | 0.09 | 0.02 |
| Profitability | -0.04 | -0.01 | 0.00 | 0.00 | -0.04 | 0.15 | 0.19 | 0.12 |

| | Labor intensity | Marginality | Asset structure | Liquidity | Profitability |
|---|---|---|---|---|---|
| Labor intensity | 1.00 | | | | |
| Marginality | 0.22 | 1.00 | | | |
| Asset structure | 0.24 | 0.31 | 1.00 | | |
| Liquidity | 0.27 | 0.28 | 0.91 | 1.00 | |
| Profitability | 0.20 | 0.61 | 0.35 | 0.33 | 1.00 |

*Table 3 – Hard innovation estimation results*

| | (1) OLS | (2) OLS | (3) 2SLS | (4) 2SLS | (5) 2SLS |
|---|---|---|---|---|---|
| | | | First stage | Second stage | Second stage |
| Dependent variables: | Patent applications | Granted patents | WoB | Patent applications | Granted patents |
| Lagged ind. variables: | | | | 0.7652** | 0.3682** |
| WoB | -0.0082 | -0.0157* | | (0.3327) | (0.1471) |
| | (0.0092) | (0.0086) | | | |
| IV CLQ | | | 0.0270*** | | |
| | | | (0.0064) | | |
| Age | -0.0148** | -0.0156*** | -0.0077 | -0.0163* | -0.0162*** |
| | (0.0060) | (0.0057) | (0.0081) | (0.0090) | (0.0059) |
| Capitalization | 0.0237 | 0.0055 | 0.0620*** | -0.0252 | -0.0185 |
| | (0.0151) | (0.0135) | (0.0112) | (0.0236) | (0.0149) |
| Labor intensity | 0.0778*** | 0.0616*** | -0.0804*** | 0.1029*** | 0.0761*** |
| | (0.0212) | (0.0189) | (0.0207) | (0.0295) | (0.0172) |
| Marginality | -0.0087* | -0.0062 | -0.0019 | -0.0100* | -0.0080 |
| | (0.0048) | (0.0044) | (0.0061) | (0.0055) | (0.0051) |
| Asset structure | -0.0037 | 0.0022 | 0.0000 | -0.0032 | 0.0025 |
| | (0.0036) | (0.0033) | (0.0030) | (0.0037) | (0.0031) |
| Liquidity | -0.0226 | -0.0246 | -0.0078 | -0.0149 | -0.0206 |
| | (0.0239) | (0.0153) | (0.0149) | (0.0249) | (0.0174) |
| Profitability | 0.0028 | -0.0023 | -0.0612*** | 0.0481 | 0.0213 |
| | (0.0143) | (0.0142) | (0.0158) | (0.0297) | (0.0163) |
| Firm fixed effects | Yes | Yes | Yes | Yes | Yes |
| Year fixed effects | Yes | Yes | Yes | Yes | Yes |
| Observations | 20812 | 20812 | 20812 | 20812 | 20812 |
| F statistic | | | 17.67 | | |
| LM statistic | | | 4.97 | | |
| Adj R-squared | 0.453 | 0.359 | 0.742 | 0.487 | 0.363 |

NUTS3 level clustered robust standard errors in parentheses. *** p<0.01, ** p<0.05, * p<0.1

*Table 4 – Soft innovation estimation results*

| | (1) OLS | (2) OLS | (3) 2SLS | (4) 2SLS | (5) 2SLS |
|---|---|---|---|---|---|
| Dependent variables: | Trademark applications | Registered trademarks | First stage WoB | Second stage Trademark applications | Second stage Registered trademarks |
| Lagged ind. variables: | | | | | |
| WoB | 0.0142** | 0.0132* | | | |
| | (0.0064) | (0.0071) | | | |
| IV CLQ | | | 0.0270*** | 0.4837** | 0.4239** |
| | | | (0.0064) | (0.1929) | (0.1771) |
| Age | -0.0138** | -0.0150** | -0.0077 | -0.0165** | -0.0175*** |
| | (0.0053) | (0.0053) | (0.0080) | (0.0065) | (0.0063) |
| Capitalization | 0.0007 | 0.0019 | 0.0620*** | -0.0266* | -0.02207 |
| | (0.0099) | (0.0101) | (0.0110) | (0.0141) | (0.0139) |
| Labor intensity | 0.0258 | 0.0300* | -0.0804*** | 0.0453 | 0.0480 |
| | (0.0172) | (0.0169) | (0.0207) | (0.0306) | (0.0291) |
| Marginality | -0.0102*** | -0.0100** | -0.0019 | -0.0095* | -0.0095* |
| | (0.0035) | (0.0039) | (0.0061) | (0.0050) | (0.0054) |
| Asset structure | 0.0018 | 0.0020 | 0.0000 | 0.0023 | 0.0026 |
| | (0.0025) | (0.0028) | (0.0030) | (0.0033) | (0.0039) |
| Liquidity | 0.0163 | 0.0168 | -0.0078 | 0.0180 | 0.0163 |
| | (0.0170) | (0.018) | (0.015) | (0.0187) | (0.0192) |
| Profitability | 0.0390** | 0.0376** | -0.0612*** | 0.0660** | 0.0616*** |
| | (0.0145) | (0.0138) | (0.0158) | (0.0255) | (0.0232) |
| Firm fixed effects | Yes | Yes | Yes | Yes | Yes |
| Year fixed effects | Yes | Yes | Yes | Yes | Yes |
| Observations | 20812 | 20812 | 20812 | 20812 | 20812 |
| F statistic | | | 17.67 | | |
| LM statistic | | | 4.97 | | |
| Adj R-squared | 0.453 | 0.359 | 0.742 | 0.357 | 0.350 |

NUTS3 level clustered robust standard errors in parentheses. *** p<0.01, ** p<0.05, * p<0.1

*Table 5 - Robustness checks estimation results*

| | Hard innovation | | Soft innovation | | | Hard innovation | | Soft Innovation | |
|---|---|---|---|---|---|---|---|---|---|
| | (1) OLS | (2) OLS | (3) OLS | (4) OLS | (5) 2SLS | (6) 2SLS | (7) 2SLS | (8) 2SLS | (9) 2SLS |
| | | | | | First stage | Second stage | Second stage | Second stage | Second stage |
| Dependent variables: | Patent applications | Granted patents | Trademark applications | Registered trademarks | WoB | Patent applications | Granted patents | Trademark applications | Registered trademarks |
| Lagged ind. variables: | -0.0078 | -0.0148* | 0.0120** | 0.0118** | | 0.6793** | 0.3267** | 0.4294** | 0.3763** |
| WoB | (0.0089) | (0.0082) | (0.0046) | (0.0048) | | (0.2935) | (0.1291) | (0.1777) | (0.1630) |
| IV CLQ | | | | | 0.0305*** | | | | |
| | | | | | (0.0073) | | | | |
| Age | -0.0149** | -0.0157*** | -0.0138** | -0.0150** | -0.0121 | -0.0140 | -0.0151*** | -0.0150** | -0.0162*** |
| | (0.0059) | (0.0054) | (0.0053) | (0.0053) | (0.0073) | (0.0085) | (0.0057) | (0.0057) | (0.0056) |
| Capitalization | 0.0237** | 0.0055 | 0.0008 | 0.0020 | 0.0613*** | -0.0193 | -0.0157 | -0.0229* | -0.0188 |
| | (0.0109) | (0.0100) | (0.0097) | (0.0100) | (0.0115) | (0.0216) | (0.0219) | (0.0132) | (0.0131) |
| Labor intensity | 0.0778*** | 0.0616*** | 0.0257 | 0.0300* | -0.0892*** | 0.1020*** | 0.0757*** | 0.0447 | 0.0475 |
| | (0.0168) | (0.0154) | (0.0173) | (0.0170) | (0.0194) | (0.0277) | (0.0220) | (0.0320) | (0.0302) |
| Marginality | -0.0087 | -0.0062 | -0.0102*** | -0.0100** | -0.0028 | -0.0096* | -0.0078 | -0.0092* | -0.0092* |
| | (0.0053) | (0.0045) | (0.0035) | (0.0039) | (0.0070) | (0.0055) | (0.0048) | (0.0050) | (0.0055) |
| Asset structure | -0.0037 | 0.0022 | 0.0018 | 0.0020 | -0.0023 | -0.0016 | 0.0032 | 0.0033 | 0.0035 |
| | (0.0030) | (0.0028) | (0.0025) | (0.0028) | (0.031) | (0.0037) | (0.0035) | (0.0032) | (0.0034) |
| Liquidity | -0.0226 | -0.0245 | 0.0162 | 0.0167 | -0.0035 | -0.0185 | -0.0223 | 0.0157 | 0.0143 |
| | (0.0165) | (0.0157) | (0.0170) | (0.0183) | (0.0150) | (0.0251) | (0.0156) | (0.0180) | (0.0186) |
| Profitability | 0.0029 | -0.0022 | 0.0388** | 0.0375** | -0.0571*** | 0.0401 | 0.0175 | 0.0609** | 0.0572*** |
| | (0.0135) | (0.0125) | (0.0145) | (0.0138) | (0.0150) | (0.0258) | (0.0235) | (0.0234) | (0.0215) |
| Firm fixed effects | Yes | Yes | Yes | Yes | Yes | Yes | Yes | Yes | Yes |
| Year fixed effects | Yes | Yes | Yes | Yes | Yes | Yes | Yes | Yes | Yes |
| Observations | 20812 | 20812 | 20812 | 20812 | 20812 | 20812 | 20812 | 20812 | 20812 |
| F statistic | | | | | 17.52 | | | | |

| | | | | | |
|---|---|---|---|---|---|
| LM statistic | 4.9 | | | | |
| Adj R-squared | 0.697 | 0.375 | 0.363 | 0.354 | 0.350 |

NUTS 3 level clustered robust standard errors in parentheses. *** p<0.01, ** p<0.05, * p<0.1